\documentclass[letterpaper, 10pt, conference]{ieeeconf}
\IEEEoverridecommandlockouts             
\usepackage{amsmath, amssymb, amsfonts}
\usepackage[colorlinks=true, citecolor=blue, linkcolor=magenta]{hyperref}
\let\labelindent\relax
\usepackage{url, footnote, enumitem}
\usepackage{caption, subcaption}
\usepackage{tikz}\usetikzlibrary{calc, positioning}
\usepackage[linesnumbered, lined, boxed]{algorithm2e}
\usepackage{graphicx, cite, textcomp, xcolor}

\usepackage{amsmath, amsfonts, amssymb, amsthm}
\allowdisplaybreaks
\theoremstyle{plain}
\newtheorem{remark}{Remark}
\newtheorem{assumption}{Assumption}
\newtheorem{definition}{Definition}

\newtheorem{theorem}{Theorem}

\newtheorem{fact}{Fact}

\title{\LARGE \bf Synthesis of Data-Driven Nonlinear State Observers using Lipschitz-Bounded Neural Networks}
\author{Wentao Tang$^{1}$
\thanks{*This work is supported by the author's faculty startup fund.}
\thanks{$^{1}$Wentao Tang is with the Department of Chemical and Biomolecular Engineering, North Carolina State University, Raleigh, NC 27695 {\tt\small wtang23@ncsu.edu}}
}

\begin{document}
\maketitle\thispagestyle{empty}\pagestyle{empty}

\begin{abstract} 
	This paper focuses on the \emph{model-free} synthesis of state observers for nonlinear autonomous systems without knowing the governing equations. Specifically, the Kazantzis-Kravaris/Luenberger (KKL) observer structure is leveraged, where the outputs are fed into a linear time-invariant (LTI) system to obtain the observer states, which can be viewed as the states nonlinearly transformed by an immersion mapping, and a neural network is used to approximate the inverse of the nonlinear immersion and estimate the states. In view of the possible existence of noises in output measurements, this work proposes to impose an upper bound on the Lipschitz constant of the neural network for robust and safe observation. A relation that bounds the generalization loss of state observation according to the Lipschitz constant, as well as the $H_2$-norm of the LTI part in the KKL observer, is established, thus reducing the model-free observer synthesis problem to that of Lipschitz-bounded neural network training, for which a direct parameterization technique is used. The proposed approach is demonstrated on a chaotic Lorenz system. 
\end{abstract}

\section{Introduction}
For nonlinear systems that arise from realistic engineering applications such as transport-reaction processes, modern control theory relies on \emph{state-space representations} for their modeling, analysis, and control \cite{kravaris1990geometric, baldea2012dynamics, rawlings2017model}. 
Recent advances in nonlinear control have highlighted the role of data-driven (machine learning) techniques in identifying governing equations or underlying dynamical structures \cite{brunton2016discovering, schoukens2019nonlinear, ren2022tutorial}, analyzing system and control-theoretic properties \cite{van2023informativity, martin2023guarantees}, and synthesizing model-free controllers \cite{hou2013model, nian2020review, tang2022data}. 
In these efforts, it is often assumed that the \emph{state} information is available for analysis or control; for example, in reinforcement learning (RL) literature, it is common to apply stochastic first-order optimization to learn a value (cost) function or $Q$ function based on temporal actions and state measurements. In many (if not most) control engineering applications, such as in chemical processes, however, it is more likely that the states are not measurable. 

\par 
Hence, for nonlinear control in a state-space framework, a \emph{state observer} is necessary, whereby the states are estimated based on input and output history \cite{kravaris2013advances}. A recent review on model-based approaches to synthesize state observers is found in Bernard, Andrieu, and Astolfi \cite{bernard2022observer}. 
A classical form of state observer for linear systems is known as Luenberger observer \cite{luenberger1966observers}, which an auxiliary linear time-invariant (LTI) system that uses the plant outputs as inputs and returns state estimates. The observer states are in fact a linear transform of the plant states \cite{simon2006optimal}. 
The idea was extended to nonlinear systems in the seminal work of Kazantzis and Kravaris \cite{kazantzis1998nonlinear}. In their Kazantzis-Kravaris/Luenberger (KKL) observer (as named in Andrieu and Praly \cite{andrieu2006existence}) still uses an LTI system to convert plant outputs to observer states, which turn out to be the plant states transformed via a nonlinear immersion. Thus, the observer synthesis problem reduces to the determination of this nonlinear immersion and its inverse, via solving (model-based) partial differential equations (PDEs). 
Such a KKL observer was extended from autonomous to actuated systems in \cite{bernard2018luenberger}, where the LTI part is replaced by an input-affine one with an additional nonlinear drift term associated with the actuated inputs. 

\par 
This paper focuses on the \emph{synthesis of KKL observer} in a \emph{model-free} manner, without assuming prior knowledge on the plant dynamics. This is motivated by two reasons: (i) many nonlinear systems that involve complex kinetic or kinematic mechanisms are often hard to model accurately, and (ii) it can be challenging to solve the associated PDEs, especially in high-dimensional state space (in fact, there may not be well-posed boundary conditions). 
In the recent years, there have been several works that pioneered the use of neural networks in the observer problem. For example, Ramos et al. \cite{ramos2020numerical} first trained neural networks to approximate the inverse immersion map to reconstruct the actual states from observer states. Then, the optimization of pole placement was considered along with the training of inverse immersion in \cite{buisson2022towards}. 
Niazi et al. \cite{niazi2022learning} used physics-informed neural networks (PINNs) to approach a surrogate solution to solve the PDEs. Miao and Gatsis \cite{miao2022learning} formulated a dynamic optimization problem to minimize the accumulated squared state observation error, whereby the optimality condition, through calculus of variations results in neural ODEs. 

\par 
It is commonly known that neural networks, when overparameterized with large widths and depths, may cause a deteriorated capability of generalization. It has also been argued that neural networks can be fragile to adversarial attacks to the training data and thus must be equipped with a self-defense mechanisms that warranty robustness \cite{huang2017adversarial, zhang2019theoretically}. 
In particular, controlling the Lipschitz constant of the mapping specified by the neural network has been studied as a promising approach \cite{fazlyab2019efficient, latorre2020lipschitz, pauli2021training}. However, in these works, estimating and minimizing the Lipschitz constant requires the use of semidefinite programming routines, which has a high complexity when the number of neurons is large. 
An alternative way, called \emph{direct paramterizaton}, as recently proposed in Wang and Manchester \cite{wang2023direct}, is to translate the Lipschitz bound constraint into a special architecture of the neural layers, thus allowing the use of typical back-propagation (BP) to train the network in an unconstrained way. 

Hence, in this work, the Wang-Manchester direct parameterization is adopted to train Lipschitz-bounded neural networks in a KKL state observer for any unknown nonlinear autonomous system. The paper establishes a relation between the generalized observation error and the Lipschitz bound of the neural network as well as the $H_2$-norm of the LTI observer dynamics, under a typical white noise assumption on the plant outputs. Hence, by varying the Lipschitz bound, the optimal observer can be synthesized.

\section{Preliminaries}
\par We consider a nonlinear autonomous system:
\begin{equation}\label{eq:system}
	\dot{x}(t) = f(x(t)), \quad y(t) = h(x(t))
\end{equation}
where $x(t)\in \mathcal{X}\subseteq \mathbb{R}^n$ is the vector of states and $y(t)\in\mathbb{R}^m$ represents the outputs. For simplicity, we will consider $m=1$. It is assumed that $f$ and $h$ are smooth on $\mathcal{X}$ to guarantee existence and uniqueness of solution but unknown for model-based synthesis.

\subsection{KKL Observer}
For nonlinear systems, KKL observer generalizes the notion of Luenberger observers that were restricted to linear systems \cite{luenberger1966observers}, providing a generic method for state observation with mild assumptions to guarantee existence. Specifically, the KKL observer for \eqref{eq:system} is expressed as
\begin{equation}\label{eq:KKL}
	\dot{z}(t) = Az(t) + By(t), \quad \hat{x}(t) = T^\dagger (z(t)). 
\end{equation}
Here the observer states $z\in\mathbb{R}^{n_z}$ has an LTI dynamics. The matrices $A$ and $B$ are chosen under the requirements of (i) controllability of $(A, B)$ should be controllable, (ii) Hurwitz property of $A$, and (iii) sufficiently high dimension of $z$ ($n_z$), which should be at least $n+1$ if $(A, B)$ is complex \cite{andrieu2006existence} and at least $2n+1$ if $(A, B)$ is real \cite{brivadis2023further}. The mapping from the observer states $z$ to the state estimates $\hat{x}$ is a static one, $T^\dagger$, which is the left-pseudoinverse of a nonlinear immersion $T$ (i.e., a differentiable injection satisfying $T^\dagger \circ T = \mathsf{id}$). This immersion $T$ should satisfy the following PDE:
\begin{equation}\label{eq:PDE}
	\frac{\partial T}{\partial x}(x)f(x) = AT(x) + Bh(x), \quad \forall x\in\mathcal{X}, 
\end{equation}
where $\partial T/\partial x$ denotes the Jacobian matrix of $T$. It can be easily verified that under the above PDE, $dT(x)/dt = AT(x) + By$, and thus $z - T(x)$ has an exponentially decaying dynamics, as $A$ is Hurwitz. 

\par The conditions for the existence of a KKL observer, namely the solution to its defining PDE \eqref{eq:PDE}, have been established based on the condition of backward distinguishability. In below, we denote the solution to the ODEs $\dot{x} = f(x)$ at time $t$ with initial condition $x(0) = \xi$ as $\Phi_t(\xi)$. For any open set $\mathcal{O}$ in $\mathcal{X}$, denote the backward time instant after which the solution does not escape this region by $\varsigma_\mathcal{O}(\xi) = \inf\{ t|\Phi_t(\xi) \in \mathcal{O}\}$. Also denote $\mathcal{O} + \epsilon := \{\xi + \eta | \xi\in\mathcal{O}, \|\eta\| < \epsilon \}$. 
\begin{definition}[Backward distinguishability]
	The system \eqref{eq:system} is $(\mathcal{O}, \epsilon)$-\emph{backward distinguishable} if for any distinct $\xi, \xi^\prime \in \mathcal{X}$ there exists a negative $t > \varsigma_{\mathcal{O} + \epsilon}(\xi) \wedge \varsigma_{\mathcal{O} + \epsilon}(\xi^\prime)$ such that $h(\Phi_t(\xi)) \neq h(\Phi_t(\xi^\prime))$. 
\end{definition}
\begin{fact}[Existence of KKL observer, cf. Brivadis et al. \cite{brivadis2023further}]
	Assume that there is an open $\mathcal{O} \subseteq \bar{\mathcal{X}}$ and a positive constant $\epsilon$ such that the system \eqref{eq:system} is $(\mathcal{O}, \epsilon)$-backward distinguishable. Then there exists a constant $\rho > 0$ such that for all but a Lebesbue-zero-measure set of $(A, B)\in\mathbb{R}^{(2n+1)\times (2n+1)} \times \mathbb{R}^{(2n+1)}$, if $A+\rho I$ Hurwitz, then there exists an immersion $T: \mathcal{O}\rightarrow \mathbb{R}^{(2n+1)}$ solving the PDEs \eqref{eq:PDE}. 
\end{fact}

\begin{figure}
	\centering
	\includegraphics[width=\columnwidth]{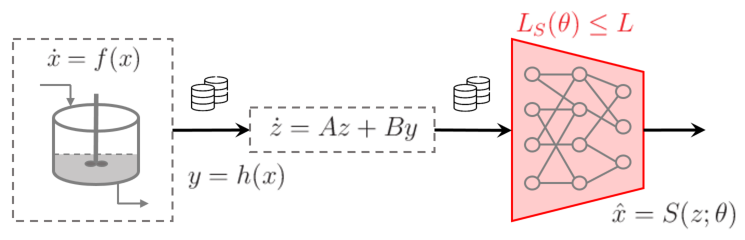}
	\caption{KKL observer with a Lipschitz-bounded neural network to be trained.}
	\label{fig:overall}
\end{figure}

The above theorem clarifies that as long as the spectrum of $A$ is restricted to the left of $-\rho+i\mathbb{R}$, the LTI dynamics in the KKL observer can be almost arbitrarily assigned. Once $(A, B)$ are chosen, the remaining question for synthesis a KKL observer \eqref{eq:KKL} is to numerically determine the solution. In view of the computational challenge in directly solving the PDEs \eqref{eq:PDE} and the recent trend of handling the problem by neural approaches \cite{ramos2020numerical, buisson2022towards, niazi2022learning}, this work will seek to approximate $T^\dagger$ by a neural network. Yet, instead of using a vanilla multi-layer perceptron architecture, a Lipschitz-bounded neural network will be adopted, which safeguards the generalization performance of state observation, which will be discussed in \S\ref{sec:analysis}. This overall idea is illustrated in Fig. \ref{fig:overall}. 

\subsection{Lipschitz-Bounded Neural Networks}
\par Consider a $\nu$-layer neural network $\hat{x} = S(z, \theta)$ with all parameters denoted as a single vector $\theta$. Without loss of generality, assume that the activation function (element-wise applied to vectors) is $\sigma: \mathbb{R}\rightarrow\mathbb{R}$, with slope bounded in $[0, 1]$ (in this work, rectified linear units (ReLU) are used to prevent gradient decay in BP training). The neural network then can be expressed as 
\begin{equation}\label{eq:neural}
	\begin{aligned}
		&z^{\ell+1} = \sigma(W^\ell z^\ell + b^\ell), \enskip \ell = 0,\dots,\nu-1 \\
		&z^0 = z, \quad \hat{x} = W^\nu z^\nu + b^\nu. 
	\end{aligned}
\end{equation}
where $W^0, \dots, W^\nu$ are the weight matrices and $b^0, \dots, b^\nu$ are the biases. In total there are $\nu$ activation layers inserted between $\nu+1$ fully connected layers. $z$ represents the inputs to the neural network and $\hat{x}$ is the output vector, as we will use such a neural network to approximate the $T^\dagger$ mapping in the KKL observer. 

\par Given a neural network with fixed parameters $\theta = (W^0, b^0, \dots, W^\nu, b^\nu)$, a rough estimate of the Lipschitz constant of $S$ can be obviously obtained as 
\begin{equation}
	L_S(\theta) = \prod_{\ell=0}^{\nu} \|W^\ell\|_2, 
\end{equation}
where $\|\cdot\|_2$ for a matrix refers to its operator norm induced by the $\ell_2$-norm of vectors, i.e., its largest singular value. To reduce the conservativeness, Fazlyab et al. \cite{fazlyab2019efficient} leverages the control-theoretic tool of integral quadratic constraints to formulate the Lipschitz bound condition as a linear matrix inequality, thus estimating the Lipschitz constants and training Lipschitz-bounded neural networks through solving semidefinite programming problems \cite{pauli2021training}. The pertinent matrix size, however, proportionally scales with the total number of neurons, which results in high computational complexity unless the neural network is very small. 

\par 
The recent work of Wang and Manchester \cite{wang2023direct} proposed a \emph{direct parameterization} approach to accommodate Lipschitz bound by a special design of the neural network architecture instead of imposing extra parameter constraints. By this approach, the training of neural networks is an unconstrained optimization problem and is thus amenable to the typical, computationally lightweight back-propagation (BP) routine. Wang-Manchester direct parameterization is conceptually related to, and arguably motivated by, the theory of controller parameterization \cite{revay2023recurrent, zheng2020equivalence}. 

\begin{figure}
	\centering
	\includegraphics[width=\columnwidth]{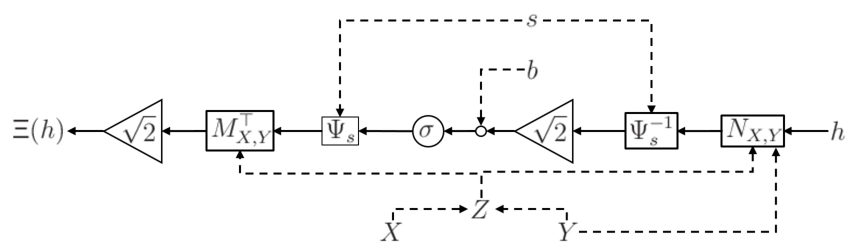}
	\caption{A sandwich layer and its parameters.}\label{fig:sandwich}
\end{figure}

\begin{definition}[$1$-Lipschitz sandwich layer, cf. \cite{wang2023direct}]
	Given parameters $X \in \mathbb{R}^{d\times d}$, $Y\in \mathbb{R}^{c\times d}$, $s\in \mathbb{R}^{d}$, and $b \in \mathbb{R}^d$, a $1$-Lipschitz sandwich layer is defined as such a mapping $\mathsf{\Xi}: \mathbb{R}^c \rightarrow \mathbb{R}^d$ that maps any $h\in\mathbb{R}^c$ into a $\mathsf{\Xi}(h; X, Y, s, b)\in \mathbb{R}^d$ according to the following formulas:
	\begin{equation}\label{eq:sandwich}
		\begin{aligned}
			Z &= X - X^\top + Y^\top Y, \enskip \Psi_s = \mathrm{diag}(e^s) \\
			M_{X, Y} &= \left[(I+Z)^{-1}(I-Z)\right]^\top, \\
			N_{X, Y} &= \left[-2Y(I+Z)^{-1}\right]^\top, \\
			\mathsf{\Xi}(h) &= \sqrt{2}M_{X, Y}^\top \Psi_s \sigma(\sqrt{2}\Psi_s^{-1} N_{X, Y}h + b). 
		\end{aligned}
	\end{equation}
\end{definition}
It turns out that the Lipschitz constant of the above-defined sandwich layer is guaranteed to be upper bounded by $1$ \cite[Theorem~3.3]{wang2023direct}. The mapping from the input $h$ to the output $\mathsf{\Xi}(h)$ can be regarded as comprising of an activation layer in the midst of two fully connected layers with related parameters. The operation from $(X, Y)$ to $(M, N)$ is known as the \emph{Cayley transform}. The structure and the parameters of a sandwich layer is shown in Fig. \ref{fig:sandwich}. 

\par Thus, by stacking a number of such sandwich layers after a scaling by $\sqrt{\gamma}$ and before a non-activated half-sandwich layer (meaning a layer containing only the terms in the parentheses of $\mathsf{\Xi}$ as in Equation \eqref{eq:sandwich}), a neural network with Lipschitz bound $\gamma$ can be obtained, for any provided $\gamma > 0$. 
\begin{definition}[Wang-Manchester network] 
	In this work, we refer to Wang-Manchester network, $S(\cdot|\theta)$, by a neural network in the following architecture: 
	\begin{equation}
		\begin{aligned}
			h^0 &= \sqrt{\gamma} z; \\
			h^{\ell + 1} &= \mathsf{\Xi}(h^\ell; X^\ell, Y^\ell, s^\ell, b^\ell), \enskip \ell = 0,1,\dots,\nu-1; \\
			\hat{x} &= \sqrt{\gamma}N_{X^\nu, Y^\nu}h^\nu + b^\nu. 
		\end{aligned}
	\end{equation}
	Here the parameters include $$\theta = \{X^\ell, Y^\ell, s^\ell, b^\ell\}_{\ell=0}^{\nu -1} \cup \{X^\nu, Y^\nu, b^\nu\} $$ which can be trained in an unconstrained way using back-propagation. The inputs and outputs of the network are $z$ and $\hat{x}$, respectively. 
\end{definition}
The above-defined Wang-Manchester network satisfies $\|S(\cdot|\theta)\|_\mathrm{Lip} \leq\gamma$. In this work, the network is defined and trained with data using PyTorch (version 2.0.1) on Google Colaboratory, which allows the auto-differentiation of a user-defined loss function with respect to the neural network parameters for the parameters to be iteratively updated.

\section{Analysis on the Generalized Loss}\label{sec:analysis}
Here we shall provide a justification for requiring a Lipschitz bound on the neural network. We will make the following standing assumptions on the training data collection procedure for subsequent analysis. 
\begin{assumption}[Ergodicity]
	Assume that a sample trajectory is collected from the system, whose initial state is sampled from a probability distribution $\mathcal{F}$ on $\mathcal{X}$. The distribution $\mathcal{F}$ is time-invariant (i.e., an eigenmeasure of the Perron-Frobenius operator), so that any point of the trajectory comes from $\mathcal{F}$. 
\end{assumption}
Suppose that The LTI dynamics of the KKL observer, $(A, B)$, is fixed. Then the observer states can be simulated from this linear dynamics. 

\begin{assumption}[Noisy measurements]
	Assume that the input signal for this LTI system is not noise-free measurements $y = h(x)$, but instead containing a white noise of unknown covariance $\sigma^2$. In other words, the simulation from $y$ to $z$ is 
	\begin{equation}
	\begin{aligned}
		& \dot{z} = Ax + By + w, \quad \mathbb{E}[w(t)] = 0, \,\forall t\in\mathbb{R} \\
		& \mathbb{E}[w(t)w(s)] = \delta(t-s)\sigma^2 , \,\forall t, s\in\mathbb{R}.
	\end{aligned}
	\end{equation}
\end{assumption}
In this way, the collected sample, denoted as $\{(x(t_i), z(t_i))\}_{i=1}^m = \{(x_i, z_i)\}_{i=1}^m$, in fact satisfies the following relation:
\begin{equation}
	z_i = \bar{z}_i + v_i, \quad \delta_i = \int_{-\infty}^{t_i} g(\tau) w(t_i - \tau)d\tau. 
\end{equation}
Here $g(\tau)$ is the impulse response of LTI system $(A, B)$;
$\bar{z}$ is the value of $z(t_i)$ that would be otherwise obtained if there were no white noises in the output measurements. 

\begin{assumption}[Sufficient decay] 
	After a significantly long time $t_\epsilon$, $\|z - T(x)\|\leq \epsilon$ for a small enough $z$. Here $T$ is the nonlinear immersion map specified by \eqref{eq:PDE}.
\end{assumption}
Then, $\|\bar{z}_i - T(x_i)\|\leq \epsilon$. Thus, we may write 
\begin{equation}
	z_i = T(x_i) + v_i + v_i^\prime, \quad \|v_i^\prime\| \leq \epsilon. 
\end{equation}

\par Now we suppose that the sample $\{(x_i, z_i)\}_{i=1}^m$ is used to train a neural network $S(\cdot|\theta)$, which gives the state observations $\hat{x}_i = S(z_i|\theta)$, and that the resulting empirical loss, if defined as the average squared observation error, is
\begin{equation}\label{eq:empirical.loss}
	\hat{R}_S(\theta) := \frac{1}{m}\sum_{i=1}^m \|\hat{x}_i - x_i\|^2. 
\end{equation}
Then we get
\begin{equation}
	\hat{R}_S(\theta) = \frac{1}{m}\sum_{i=1}^m \left\|S\left(T(x_i) + v_i + v_i^\prime|\theta \right) - x_i \right\|^2. 
\end{equation}

\begin{assumption}
	Assume that the probability distribution $\mathcal{F}$ is supported by a compact set, i.e., if $x\sim \mathcal{F}$, then $x$ should be almost surely bounded. 
\end{assumption}
It follows that both $S(\cdot|\theta)$ and $T$ should be Lipschitz continuous. Denote their Lipschitz constants as $L_S(\theta)$ and $L_T$, respectively. We have 
\begin{equation}
	\| S\left(T(x_i) + \delta_i + \delta_i^\prime|\theta \right) - S\left(T(x_i)|\theta \right) \| \leq L_S(\theta) L_T (\|v_i\| + \epsilon). 
\end{equation}
Denote $D$ as the essential upper bound of $\|x\|$ on the distribution $\mathcal{F}$. As such, without loss of generality, let $S(T(0)) = 0$. Then $\|x - S(T(x))\|\leq (L_S(\theta)L_T + 1)D$ almost surely. Combining the above two equations, we further get
\begin{equation}
	\begin{aligned}
		& \frac{1}{m}\sum_{i=1}^m \|x_i - S(T(x_i)|\theta)\|^2 \leq \hat{R}_S(\theta) \\
		& \quad + \frac{1}{m}\sum_{i=1}^m  L_S(\theta)L_T(L_S(\theta)L_T + 1)D (\|v_i\| + \epsilon) \\
		& \quad + \frac{1}{m}\sum_{i=1}^m  L_S^2 (\theta)L_T^2 (\|v_i\| + \epsilon)^2. 
	\end{aligned}
\end{equation}
That is, 
\begin{equation}
	\begin{aligned}
		& \frac{1}{m}\sum_{i=1}^m \|x_i - S(T(x_i)|\theta)\|^2 \leq \hat{R}_S(\theta) \\
		& \quad + \frac{1}{m}\sum_{i=1}^m  (L_S(\theta)L_T + 1)^2 \left(D + \|v_i\| + \epsilon \right)(\|v_i\| + \epsilon).
	\end{aligned}
\end{equation}
The left-hand side gives an estimation of the empirical loss when observing the states from perfect output measurements (namely when $z_i = T(x_i)$, $i=1,\dots,m$). Further expanding the last term and applying Cauchy-Schwarz inequality, we have
\begin{equation}
	\begin{aligned}
		& \frac{1}{m}\sum_{i=1}^m \|x_i - S(T(x_i)|\theta)\|^2 \leq \hat{R}_S(\theta) + (L_S(\theta)L_T + 1)^2 \times \\
		& \quad \left[ \frac{1}{m}\sum_{i=1}^m \|v_i\|^2 + (D + 2\epsilon)\left(\sum_{i=1}^m \|v_i\|^2\right)^{1/2} + (D+\epsilon)\epsilon \right].
	\end{aligned}
\end{equation}

\par Given that $v_i$ is the response of LTI system $(A, B)$ to a white noise of covariance $\sigma^2$, $\mathbb{E}(\|v_i\|^2) = h^2\sigma^2$ where $h$ is the $H_2$-norm of the system $(A, B)$ where $A$ is Hurwitz. Therefore, 
\begin{equation}
	\mathbb{E}\left(\frac{1}{m}\sum_{i=1}^m \|v_i\|^2\right) = h^2\sigma^2.
\end{equation}
Let $\alpha$ be a small positive number. With confidence $1-\alpha/2$, a  conservative estimation for its upper bound can be found according to Markov inequality:
\begin{equation}
	\frac{1}{m}\sum_{i=1}^m \|v_i\|^2 \leq \frac{1}{1-\alpha/2} h^2\sigma^2. 
\end{equation}
Therefore, 
\begin{equation}\label{eq:generalization.1}
	\begin{aligned}
		& \frac{1}{m}\sum_{i=1}^m \|x_i - S(T(x_i)|\theta)\|^2 \leq \hat{R}_S(\theta) + (L_S(\theta)L_T + 1)^2 \times \\
		& \quad \left[ \frac{h^2\sigma^2}{1-\alpha/2}  + (D + 2\epsilon)\frac{h\sigma}{\sqrt{1-\alpha/2}} + (D+\epsilon)\epsilon \right].
	\end{aligned}
\end{equation}

\par Finally, we note that for $x\sim \mathcal{F}$, now that $\|x - S(T(x)|\theta)\|\leq (L_S(\theta)L_T + 1)D$ almost surely, by Hoeffding's inequality, for any $\varepsilon > 0$, 
\begin{equation}
	\begin{aligned}
		& \mathbb{P}\bigg(\bigg{|} \frac{1}{m}\sum_{i=1}^m \|x_i - S(T(x_i)|\theta)\|^2 - \mathbb{E}\left(\|x - S(T(x))\|^2\right) \bigg{|} \\
		& \geq (L_S(\theta)L_T + 1)^2D^2 \varepsilon \bigg) \leq 2\exp\left(-2m\varepsilon^2\right).
	\end{aligned}
\end{equation}
Thus, with confidence $1-\alpha/2$, we have 
\begin{equation}\label{eq:generalization.2}
	\begin{aligned}
		& \bigg{|} \frac{1}{m}\sum_{i=1}^m \|x_i - S(T(x_i)|\theta)\|^2 - \mathbb{E}\left(\|x - S(T(x)|\theta)\|^2\right) \bigg{|} \\
		& < (L_S(\theta)L_T + 1)^2D^2 \sqrt{\frac{\ln(4/\alpha)}{2m}}.
	\end{aligned}
\end{equation}
Combining \eqref{eq:generalization.1} and \eqref{eq:generalization.2}, we have the following theorem. 
\begin{theorem}
	Under the afore-mentioned assumptions, the generalization loss, defined as \begin{equation}
		R_S(\theta) = \mathbb{E}\left(\|x - S(T(x)|\theta)\|^2\right), 
	\end{equation}
	is related to the empirical loss as defined in \eqref{eq:empirical.loss} by 
	\begin{equation}
		R_S(\theta) < \hat{R}_S(\theta) + (L_S(\theta)L_T + 1)^2 \Delta(h, \sigma, \alpha, \epsilon). 
	\end{equation} 
	with confidence $1 - \alpha$ ($\alpha \in (0,1)$). Here
	\begin{equation}\label{eq:Delta.term}
		\begin{aligned}
			\Delta(h, \sigma, \alpha, \epsilon) =& D^2 \sqrt{\frac{\ln(4/\alpha)}{2m}} + \frac{h^2\sigma^2}{1-\alpha/2} \\
			&+ (D + 2\epsilon)\frac{h\sigma}{\sqrt{1-\alpha/2}} + (D+\epsilon)\epsilon.
		\end{aligned}
	\end{equation}
\end{theorem}
The theorem shows that the Lipschitz constant of the neural network trained plays an important role in the generalized performance of the resulting state observer. The effect of $L_S(\theta)$ is mainly that of amplifying the first and third terms defined on the right-hand side of \eqref{eq:Delta.term}, supposing that $\sigma$ and $\epsilon$ are small enough. These two terms respectively arise from (i) the overall upper bound of the observation error $\|x - S(T(x)|\theta)\|$, which acts as a coefficient before the Hoeffding term $\sqrt{\ln(4/\alpha)/2m}$, and (ii) the effect of noisy measurements on the observer states. 

\begin{remark}
	It is noted that the performance bound stated in the above theorem can be conservative. The conclusion that $L_S(\theta)$ amplifies the generalization error and measurement noise should be considered as qualitative. The theorem also does not suggest a tractable algorithm to optimize the selection of $(A, B)$ along with the neural network $S(\cdot|\theta)$, as the dependence of $L_T$ on $(A, B)$ is highly implicit. Hence, this paper does not consider the problem of simultaneously training $(A, B)$ and the neural network. 
\end{remark}

\section{Case Study}\label{sec:examples}
Let us consider a Lorenz system in a 3-dimensional state space with chaotic behavior. The equation is written as:
\begin{equation}
	\begin{aligned}
	\dot{x}_1 &= 10(x_2-x_1), \\
	\dot{x}_2 &= x_1(28-10x_3) - x_2, \\
	\dot{x}_3 &= 10x_1x_2 - (8/3)x_3.
	\end{aligned}
\end{equation}
Suppose that the measurement used for state observation is $y = x_2$, where a white noise exists. We assign different values to the variance of the measurement noise and investigate how the resulting neural network should be chosen differently. To simulate the process we will use a sampling time of $0.01$. The LTI part of the KKL observer, $A = -\mathrm{diag}(8, 4, 2, 1)$ and $B = [1, 1, 1, 1]^\top$ are chosen. At the beginning of the observer simulation, $z(0) = 0$ is set as the initial condition; we simulate the dynamics until $t = 500$ and randomly collect $m = 2000$ time instants between $t = 20$ and $t=500$ as the training data. 

\par
\begin{figure}
	\centering
	\begin{subfigure}[b]{\columnwidth}
		\centering
		\includegraphics[width=0.9\columnwidth]{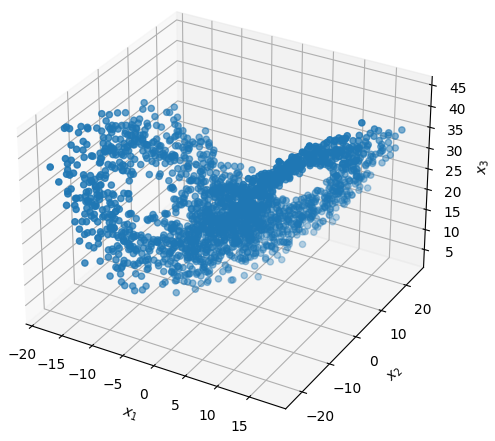}
		\caption{$x$ sample. }\label{fig:sample-x}
	\end{subfigure}
	\hfill
	\begin{subfigure}[b]{\columnwidth}
		\centering
		\includegraphics[width=0.9\columnwidth]{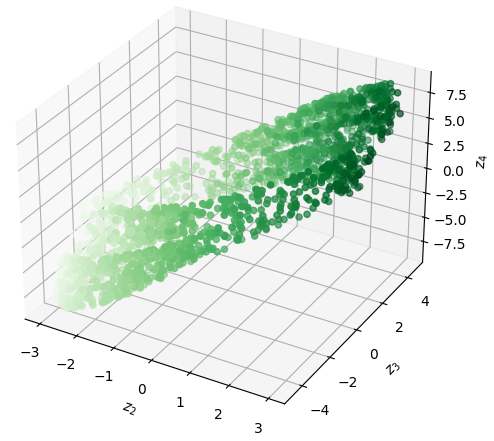}
		\caption{$z$ sample. (The $z_1$ component is reflected by the marker color.)}\label{fig:sample-z}
	\end{subfigure}
	\caption{Sample collected from the Lorenz system.}
	\label{fig:sample}
\end{figure}

 Consider first the case with noiseless measurement ($\sigma = 0$). The sample $\{(x_i, z_i)\}_{i=1}^{2000}$ is plotted in Fig. \ref{fig:sample}, which shows that the data points are representative on the forward invariant set of the system, and that the observer states $z_i$ indeed captures the structure of such a Lorenz attractor in a $4$-dimensional space. Hence, we train the Wang-Manchester network using a randomly selected $80\%$ of the sample under the mean-squares loss metric, and validate using the remaining $20\%$ sample points. Stochastic gradient descent (SGD) algorithm with a learning rate of $10^{-3}$ is used for optimization. The number of epochs is empirically tuned to $300$. The neural network has $2$ hidden layers, each containing $8$ neurons, resulting in $292$ parameters to train in total. After training, the Lipschitz constant is evaluated a posteriori via the semidefinite programming approach of Fazlyab et al. \cite{fazlyab2019efficient} using \texttt{cvxpy}, which costs approximately $1.5$ seconds (for a randomly initialized network). 

\par
\begin{figure}
	\centering
	\begin{subfigure}[b]{\columnwidth}
		\centering
		\includegraphics[width=0.9\columnwidth]{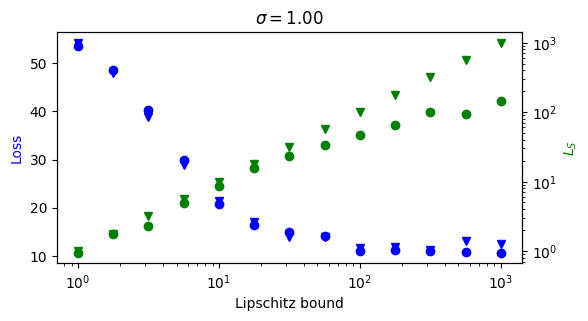}
	\end{subfigure}
	\hfill
	\begin{subfigure}[b]{\columnwidth}
		\centering
		\includegraphics[width=0.9\columnwidth]{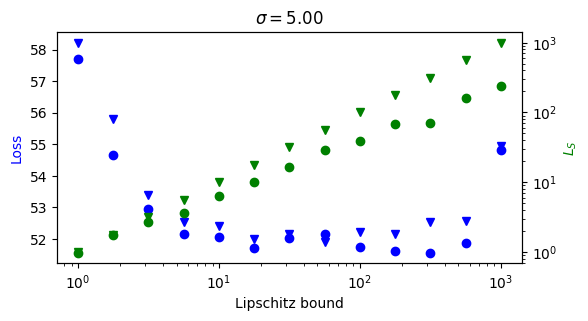}
	\end{subfigure}
	\hfill
	\begin{subfigure}[b]{\columnwidth}
		\centering
		\includegraphics[width=0.9\columnwidth]{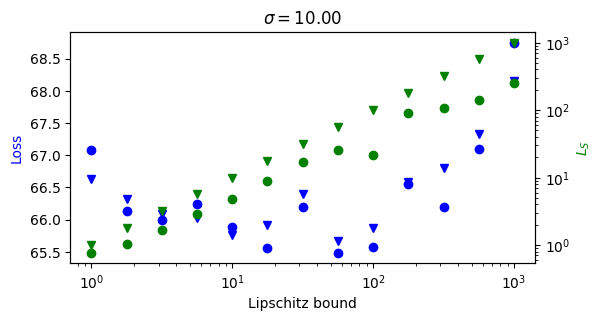}
	\end{subfigure}
	\caption{Loss and Lipschitz constants under different prior Lipschitz bounds. (Blue wedges: training loss, blue circles: validation loss, green circles: prior Lipschitz bound; green wedges: posterior Lipschitz bound.)}
	\label{fig:training}
\end{figure}

 Varying the prior bound on the Lipschitz constant, the resulting training loss, validation loss, and the posterior Lipschitz bound obtained from the same training conditions are illustrated in Fig. \ref{fig:training}. The following observations can be made from these results. 
\begin{itemize}
	\item As anticipated, as the set bound on the Lipschitz bound increases, the Lipschitz constant of the trained neural network becomes higher. The Lipschitz constants estimated a posteriori are lower than the prior bound on the Wang-Manchester network, validating the direct parameterization approach on constraining the slope. On the other hand, the actually posterior Lipschitz constant has an increasingly large lag behind the prior bound; for example, when the prior bound is $1000$, the $L_S$ after training does not exceed $300$. This indicates that even for the training objective alone, there is a ``resistance'' to pursue the maximally possible Lipschitz constant. 
	\item When the Lipschitz bound is small, relaxing the restriction on $L_S$ is beneficial for decreasing the training loss as well as the validation loss, showing that the Lipschitz bound is a bottleneck causing underfitting. When $L_S$ is high enough, such underfitting no longer exists; instead, overfitting will appear, with rising training and validation losses. The overfitting phenomenon is more significant when the noise is large. Thus, there should be optimal values to be set as the Lipschitz bound. 
	\item Depending on the noise magnitude, the deviation of posterior Lipschitz constant from the prior bound and the emergence of overfitting phenomenon occur at different threshold values of the Lipschitz bound. Thus, the Lipschitz bound to be used for neural network training should be tuned differently as the noise intensity varies. For example, at $\sigma=1$, a suitable choice can be $\gamma=100$, whereas at $\sigma=5$ and $\sigma=10$, $\gamma$ can be chosen as $30$ and $10$, respectively. 
\end{itemize}

\par
\begin{figure}
	\centering
	\includegraphics[width=\columnwidth]{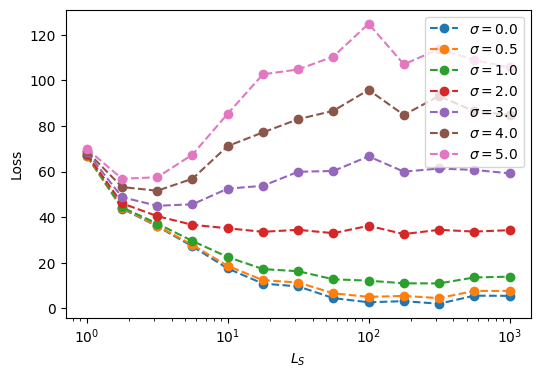}
	\caption{Errors of noiselessly trained observers in noisy environments.}
	\label{fig:generalization}
\end{figure}

 Now suppose that at the observer design stage, the Wang-Manchester network is trained by the simulated data from a perfect digital twin of the true dynamics, i.e., $\sigma=0$; yet, when applying the network trained to observe the states of the physical system, the environment is noisy. In Fig. \ref{fig:generalization}, the resulting loss (mean squared state observation error) is plotted against varying prior Lipschitz bounds under multiple values of the environment noise magnitude. It is seen that when the noise is low, roughly speaking, increasing $L_S$ leads to monotonic decrease in the observation error within a large range. On the other hand, when the environment is highly noisy (e.g., when $\sigma \geq 3$), the Lipschitz bound has a severe effect on the generalization loss, and since the achievable performance is restrictive, the fine-tuning of Lipschitz bound as a hyperparameter becomes critical. 

\par
\begin{figure}
	\centering
	\includegraphics[width=\columnwidth]{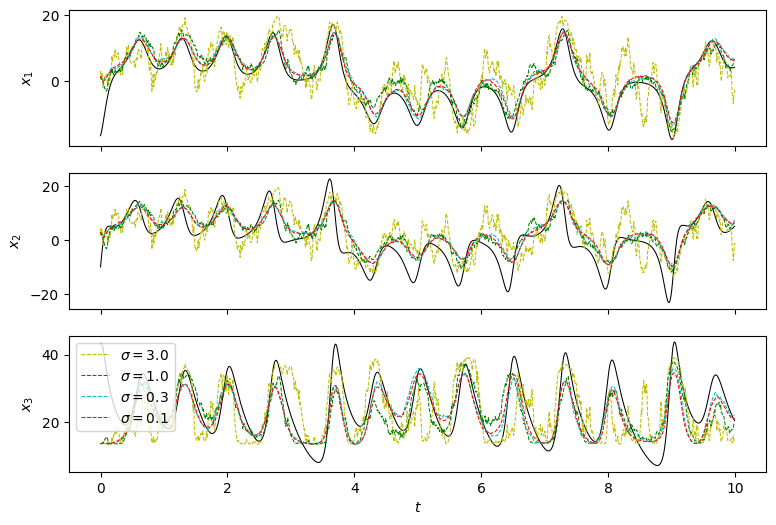}
	\caption{Estimated states by the observers in noisy environments.}
	\label{fig:simulation}
\end{figure}

 Finally, the performance of the state observer is examined. Consider using the network trained with noiseless simulation data under the prior Lipschitz bound $L_S=10$, and applying it to environments with noise $\sigma = 0.1$, $0.3$, $1.0$, $3.0$. The trajectories of the three components of estimated states by the observer are plotted against the true states in Fig. \ref{fig:simulation}, within a time horizon of $10$ time units. Naturally, when $\sigma$ is low, the state estimates can well track the true states and capture the trends in the correct directions; as $\sigma$ increases, the accuracy is lowered and the signals constructed by the observer are more noisy, occasionally yielding incorrect directions of evolution (e.g., on $3 < t < 4$ or $8 < t < 9$, where the states swing between the two foils of the Lorenz attractor). Overall, the state estimates mollifies the true state trajectories, which is due to the structure of our KKL observer -- a linear filter (LTI system) as the state dynamics and a Lipschitz-bounded neural network as the static output map. 

\section{Conclusions and Discussions}\label{sec:conclusion}
This work leverages the recent tools of Lipschitz-bounded neural networks for the synthesis of nonlinear state observers in a model-free setting. The observer, which has a Kazantzis-Kravaris structure, turns out to have a provable generalization performance that is related to the Lipschitz constant of the trained neural network (which represents the mapping from the observer states to the plant states). As such, by varying the Lipschitz bound and re-training the neural network, the optimal training result can yield the minimum generalized state observation error. 
The importance of bounding the Lipschitz constant has been demonstrated by a numerical case study on the Lorenz system. 

\par We implicitly assumed here that a simulator of the dynamics is available, so that the true states' trajectories can be used to train the neural network. However, such ground truth for supervised learning may not actually exist in real applications, i.e., only inputs and outputs are recorded, yet a state observation mechanism is still needed or desired for feedback control. To this end, the author's recent work \cite{tang2023data} proposed a data-driven KKL observer by appending a kernel dimensionality reduction scheme to the LTI dynamics, thus obtaining estimates that are diffeomorphic to the states. 

\par Also, the current approach is yet restricted to autonomous systems. For control purposes, it should be further extended to non-autonomous ones, where the Bernard-Andrieu observer structure \cite{bernard2018luenberger} is anticipated. Also, the application of such data-driven state observers to learning control-relevant properties of nonlinear dynamical systems and controller synthesis \cite{tang2019dissipativity, tang2021dissipativity} is undergoing active research.

\bibliographystyle{ieeetran}
\bibliography{ieeeconf.bib}

\begin{thebibliography}{10}
\providecommand{\url}[1]{#1}
\csname url@rmstyle\endcsname
\providecommand{\newblock}{\relax}
\providecommand{\bibinfo}[2]{#2}
\providecommand\BIBentrySTDinterwordspacing{\spaceskip=0pt\relax}
\providecommand\BIBentryALTinterwordstretchfactor{4}
\providecommand\BIBentryALTinterwordspacing{\spaceskip=\fontdimen2\font plus
\BIBentryALTinterwordstretchfactor\fontdimen3\font minus
  \fontdimen4\font\relax}
\providecommand\BIBforeignlanguage[2]{{%
\expandafter\ifx\csname l@#1\endcsname\relax
\typeout{** WARNING: IEEEtran.bst: No hyphenation pattern has been}%
\typeout{** loaded for the language `#1'. Using the pattern for}%
\typeout{** the default language instead.}%
\else
\language=\csname l@#1\endcsname
\fi
#2}}

\bibitem{kravaris1990geometric}
C.~Kravaris and J.~C. Kantor, ``Geometric methods for nonlinear process
  control. 1. background,'' \emph{Ind. Eng. Chem. Res.}, vol.~29, no.~12, pp.
  2295--2310, 1990.

\bibitem{baldea2012dynamics}
M.~Baldea and P.~Daoutidis, \emph{Dynamics and nonlinear control of integrated
  process systems}.\hskip 1em plus 0.5em minus 0.4em\relax Cambridge University
  Press, 2012.

\bibitem{rawlings2017model}
J.~B. Rawlings, D.~Q. Mayne, and M.~Diehl, \emph{Model predictive control:
  Theory, computation, and design}, 2nd~ed.\hskip 1em plus 0.5em minus
  0.4em\relax Nob Hill, 2017.

\bibitem{brunton2016discovering}
S.~L. Brunton, J.~L. Proctor, and J.~N. Kutz, ``Discovering governing equations
  from data by sparse identification of nonlinear dynamical systems,''
  \emph{Proc. Natl. Acad. Sci. U.S.A.}, vol. 113, no.~15, pp. 3932--3937, 2016.

\bibitem{schoukens2019nonlinear}
J.~Schoukens and L.~Ljung, ``Nonlinear system identification: A user-oriented
  road map,'' \emph{IEEE Control Syst. Mag.}, vol.~39, no.~6, pp. 28--99, 2019.

\bibitem{ren2022tutorial}
Y.~M. Ren, M.~S. Alhajeri, J.~Luo, S.~Chen, F.~Abdullah, Z.~Wu, and P.~D.
  Christofides, ``A tutorial review of neural network modeling approaches for
  model predictive control,'' \emph{Comput. Chem. Eng.}, vol. 165, p. 107956,
  2022.

\bibitem{van2023informativity}
H.~J. van Waarde, J.~Eising, M.~K. Camlibel, and H.~L. Trentelman, ``The
  informativity approach to data-driven analysis and control,'' \emph{arXiv
  preprint arXiv:2302.10488}, 2023.

\bibitem{martin2023guarantees}
T.~Martin, T.~B. Sch{\"o}n, and F.~Allg{\"o}wer, ``Guarantees for data-driven
  control of nonlinear systems using semidefinite programming: A survey,''
  \emph{arXiv preprint arXiv:2306.16042}, 2023.

\bibitem{hou2013model}
Z.-S. Hou and Z.~Wang, ``From model-based control to data-driven control:
  Survey, classification and perspective,'' \emph{Inform. Sci.}, vol. 235, pp.
  3--35, 2013.

\bibitem{nian2020review}
R.~Nian, J.~Liu, and B.~Huang, ``A review on reinforcement learning:
  Introduction and applications in industrial process control,'' \emph{Comput.
  Chem. Eng.}, vol. 139, p. 106886, 2020.

\bibitem{tang2022data}
W.~Tang and P.~Daoutidis, ``Data-driven control: Overview and perspectives,''
  in \emph{American Control Conference (ACC)}.\hskip 1em plus 0.5em minus
  0.4em\relax IEEE, 2022, pp. 1048--1064.

\bibitem{kravaris2013advances}
C.~Kravaris, J.~Hahn, and Y.~Chu, ``Advances and selected recent developments
  in state and parameter estimation,'' \emph{Comput. Chem. Eng.}, vol.~51, pp.
  111--123, 2013.

\bibitem{bernard2022observer}
P.~Bernard, V.~Andrieu, and D.~Astolfi, ``Observer design for continuous-time
  dynamical systems,'' \emph{Annu. Rev. Control}, 2022.

\bibitem{luenberger1966observers}
D.~Luenberger, ``Observers for multivariable systems,'' \emph{IEEE Trans.
  Autom. Control}, vol.~11, no.~2, pp. 190--197, 1966.

\bibitem{simon2006optimal}
D.~Simon, \emph{Optimal state estimation: {Kalman}, {$H_\infty$}, and nonlinear
  approaches}.\hskip 1em plus 0.5em minus 0.4em\relax John Wiley \& Sons, 2006.

\bibitem{kazantzis1998nonlinear}
N.~Kazantzis and C.~Kravaris, ``Nonlinear observer design using {Lyapunov's}
  auxiliary theorem,'' \emph{Syst. Control Lett.}, vol.~34, no.~5, pp.
  241--247, 1998.

\bibitem{andrieu2006existence}
V.~Andrieu and L.~Praly, ``On the existence of a
  {Kazantzis-Kravaris/Luenberger} observer,'' \emph{SIAM J. Control Optim.},
  vol.~45, no.~2, pp. 432--456, 2006.

\bibitem{bernard2018luenberger}
P.~Bernard and V.~Andrieu, ``Luenberger observers for nonautonomous nonlinear
  systems,'' \emph{IEEE Trans. Autom. Control}, vol.~64, no.~1, pp. 270--281,
  2018.

\bibitem{ramos2020numerical}
L.~C. Ramos, F.~Di~Meglio, V.~Morgenthaler, L.~F.~F. da~Silva, and P.~Bernard,
  ``Numerical design of {Luenberger} observers for nonlinear systems,'' in
  \emph{59\textsuperscript{th} Conference on Decision and Control (CDC)}.\hskip
  1em plus 0.5em minus 0.4em\relax IEEE, 2020, pp. 5435--5442.

\bibitem{buisson2022towards}
M.~Buisson-Fenet, L.~Bahr, and F.~Di~Meglio, ``Towards gain tuning for
  numerical {KKL} observers,'' \emph{arXiv preprint}, 2022, {arXiv:2204.00318}.

\bibitem{niazi2022learning}
M.~U.~B. Niazi, J.~Cao, X.~Sun, A.~Das, and K.~H. Johansson, ``Learning-based
  design of {Luenberger} observers for autonomous nonlinear systems,''
  \emph{arXiv preprint}, 2022, {arXiv:2210.01476}.

\bibitem{miao2022learning}
K.~Miao and K.~Gatsis, ``Learning robust state observers using neural {ODEs},''
  \emph{arXiv preprint}, 2022, {arXiv:2212.00866}.

\bibitem{huang2017adversarial}
S.~Huang, N.~Papernot, I.~Goodfellow, Y.~Duan, and P.~Abbeel, ``Adversarial
  attacks on neural network policies,'' \emph{arXiv preprint arXiv:1702.02284},
  2017.

\bibitem{zhang2019theoretically}
H.~Zhang, Y.~Yu, J.~Jiao, E.~Xing, L.~El~Ghaoui, and M.~Jordan, ``Theoretically
  principled trade-off between robustness and accuracy,'' in \emph{Int. Conf.
  Mach. Learn.}\hskip 1em plus 0.5em minus 0.4em\relax PMLR, 2019, pp.
  7472--7482.

\bibitem{fazlyab2019efficient}
M.~Fazlyab, A.~Robey, H.~Hassani, M.~Morari, and G.~Pappas, ``Efficient and
  accurate estimation of {Lipschitz} constants for deep neural networks,''
  \emph{Advances in Neural Information Processing Systems}, vol.~32, 2019.

\bibitem{latorre2020lipschitz}
F.~Latorre, P.~Rolland, and V.~Cevher, ``Lipschitz constant estimation of
  neural networks via sparse polynomial optimization,'' \emph{arXiv preprint
  arXiv:2004.08688}, 2020.

\bibitem{pauli2021training}
P.~Pauli, A.~Koch, J.~Berberich, P.~Kohler, and F.~Allg{\"o}wer, ``Training
  robust neural networks using lipschitz bounds,'' \emph{IEEE Control Systems
  Lett.}, vol.~6, pp. 121--126, 2021.

\bibitem{wang2023direct}
R.~Wang and I.~Manchester, ``Direct parameterization of lipschitz-bounded deep
  networks,'' in \emph{International Conference on Machine Learning}.\hskip 1em
  plus 0.5em minus 0.4em\relax PMLR, 2023, pp. 36\,093--36\,110.

\bibitem{brivadis2023further}
L.~Brivadis, V.~Andrieu, P.~Bernard, and U.~Serres, ``Further remarks on {KKL}
  observers,'' \emph{Syst. Control Lett.}, vol. 172, p. 105429, 2023.

\bibitem{revay2023recurrent}
M.~Revay, R.~Wang, and I.~R. Manchester, ``Recurrent equilibrium networks:
  Flexible dynamic models with guaranteed stability and robustness,''
  \emph{IEEE Trans. Autom. Control}, 2023, in press.

\bibitem{zheng2020equivalence}
Y.~Zheng, L.~Furieri, A.~Papachristodoulou, N.~Li, and M.~Kamgarpour, ``On the
  equivalence of youla, system-level, and input--output parameterizations,''
  \emph{IEEE Trans. Autom. Control}, vol.~66, no.~1, pp. 413--420, 2020.

\bibitem{tang2023data}
W.~Tang, ``Data-driven state observation for nonlinear systems based on online
  learning,'' \emph{AIChE J.}, vol.~69, no. (in press), p. e18224, 2023.

\bibitem{tang2019dissipativity}
W.~Tang and P.~Daoutidis, ``Dissipativity learning control ({DLC}): A framework
  of input--output data-driven control,'' \emph{Comput. Chem. Eng.}, vol. 130,
  p. 106576, 2019.

\bibitem{tang2021dissipativity}
------, ``Dissipativity learning control ({DLC}): Theoretical foundations of
  input--output data-driven model-free process control,'' \emph{Syst. Control
  Lett.}, vol. 147, p. 104831, 2021.

\end{thebibliography}

\end{document}